\documentclass[twocolumn,draft,showkeys,showpacs,eqsecnum,nofootinbib,aps]{revtex4}
\renewcommand{\theequation}{\arabic{equation}}
\def\beq{\begin{equation}}
\def\eeq{\end{equation}}
\def\bea{\begin{eqnarray}}
\def\eea{\end{eqnarray}}\def\nn{\nonumber}

\def\na{\nabla}
\def\pa{\partial}

\def\nn{\nonumber}

\begin{document}
\title{Stringy Jacobi fields in Morse theory}
\author{Yong Seung Cho}
\email{yescho@ewha.ac.kr} \affiliation{National Institute for
Mathematical Sciences, 385-16 Doryong, Yuseong, Daejeon 305-340
Korea}\affiliation{Department of Mathematics, Ewha Womans
University, Seoul 120-750 Korea}
\author{Soon-Tae Hong}
\email{soonhong@ewha.ac.kr} \affiliation{Department of Science
Education and Research Institute for Basic Sciences, Ewha Womans
University, Seoul 120-750 Korea}
\date{\today}
\begin{abstract}
We consider the variation of the surface spanned by closed strings
in a spacetime manifold.  Using the Nambu-Goto string action, we
induce the geodesic surface equation, the geodesic surface
deviation equation which yields a Jacobi field, and we define the
index form of a geodesic surface as in the case of point particles
to discuss conjugate strings on the geodesic surface.
\end{abstract}
\pacs{02.40.-k, 04.20.-q, 04.90.+e, 11.25.-w, 11.40.-q}
\keywords{Nambu-Goto string action, geodesic surface, Jacobi
field, index of geodesic surface, conjugate strings} \maketitle

\section{Introduction}
\setcounter{equation}{0}
\renewcommand{\theequation}{\arabic{section}.\arabic{equation}}

It is well known that string theory~\cite{witten87,pol98} is one
of the best candidates for a consistent quantum theory of gravity
to yield a unification theory of all the four basic forces in
nature. In D-brane models~\cite{pol98}, closed strings represent
gravitons propagating on a curved manifold, while open strings
describe gauge bosons such as photons, or matter attached on the
D-branes. Moreover, because the two ends of an open string can
always meet and connect, forming a closed string, there are no
string theories without closed strings.

On the other hand, the supersymmetric quantum mechanics has been
exploited by Witten~\cite{witten82} to discuss the Morse
inequalities~\cite{morse34,milnor63,wald84}. The Morse indices for
pair of critical points of the symplectic action function have
been also investigated based on the spectral flow of the Hessian
of the symplectic function~\cite{floer}, and on the Hilbert spaces
the Morse homology~\cite{schwarz} has been considered to discuss
the critical points associated with the Morse index~\cite{majer}.
The string topology was initiated in the seminal work of Chas and
Sullivan~\cite{chas99}. Using the Morse theoretic techniques,
Cohen in Ref.~\cite{biran04} constructs string topology operations
on the loop space of a manifold and relates the string topology
operations to the counting of pseudo-holomorphic curves in the
cotangent bundle.  He also speculates the relation between the
Gromov-Witten invariant~\cite{mcduff94} of the cotangent bundle
and the string topology of the underlying manifold.  Recently, the
Jacobi fields and their eigenvalues of the Sturm-Liouville
operator associated with the particle geodesics on a curved
manifold  have been investigated~\cite{hong03}, to relate the
phase factor of the partition function to the eta invariant of
Atiyah~\cite{atiyah75,witten89}.

In this paper, we will exploit the Nambu-Goto string action to
investigate the geodesic surface equation and the geodesic surface
deviation equation associated with a Jacobi field.  The index form
of a geodesic surface will be also discussed for the closed
strings on the curved manifold.

In Section II, the string action will be introduced to investigate
the geodesic surface equation in terms of the world sheet currents
associated with $\tau$ and $\sigma$ world sheet coordinate
directions. By taking the second variation of the surface spanned
by closed strings, the geodesic surface deviation equation will be
discussed for the closed strings on the curved manifold. In
Section III, exploiting the orthonormal gauge, the index form of a
geodesic surface will be also investigated together with breaks on
the string tubes. The geodesic surface deviation equation in the
orthonormal gauge will be exploited to discuss the Jacobi field on
the geodesic surface.

\section{Stringy geodesic surfaces in Morse theory}
\setcounter{equation}{0}
\renewcommand{\theequation}{\arabic{section}.\arabic{equation}}

In analogy of the relativistic action of a point particle, the
action for a string is proportional to the area of the surface
spanned in spacetime manifold $M$ by the evolution of the string.
In order to define the action on the curved manifold, let $(M,
g_{ab})$ be a $n$-dimensional manifold associated with the metric
$g_{ab}$.  Given $g_{ab}$, we can have a unique covariant
derivative $\na_{a}$ satisfying~\cite{wald84} $\na_{a}g_{bc}=0$,
$\na_{a}\omega^{b}=\pa_{a}\omega^{b}+\Gamma^{b}_{~ac}~\omega^{c}$
and \beq
(\na_{a}\na_{b}-\na_{b}\na_{a})\omega_{c}=R_{abc}^{~~~d}~\omega_{d}.\label{rtensor}
\eeq

We parameterize the closed string by two world sheet coordinates
$\tau$ and $\sigma$, and then we have the corresponding vector
fields $\xi^{a}=(\pa/\pa\tau)^{a}$ and
$\zeta^{a}=(\pa/\pa\sigma)^{a}$.  The Nambu-Goto string action is
then given by~\cite{nambu70,witten87,pol98} \beq S=-\int\int~d\tau
d\sigma f(\tau,\sigma)\eeq where the coordinates $\tau$ and
$\sigma$ have ranges $0\leq \tau\leq T$ and $0\leq \sigma\leq
2\pi$ respectively and \beq
f(\tau,\sigma)=[(\xi\cdot\zeta)^{2}-(\xi\cdot\xi)(\zeta\cdot\zeta)]^{1/2}.\eeq
We now perform an infinitesimal variation of the tubes
$\gamma_{\alpha}(\tau,\sigma)$ traced by the closed string during
its evolution in order to find the geodesic surface equation from
the least action principle.  Here we impose the restriction that
the length of the string circumference is $\tau$ independent. Let
the vector field $\eta^{a}=(\pa/\pa \alpha)^{a}$ be the deviation
vector which represents the displacement to an infinitesimally
nearby tube, and let $\Sigma$ denote the three-dimensional
submanifold spanned by the tubes $\gamma_{\alpha}(\tau,\sigma)$.
We then may choose $\tau$, $\sigma$ and $\alpha$ as coordinates of
$\Sigma$ to yield the commutator relations, \bea
\pounds_{\xi}\eta^{a}&=&\xi^{b}\na_{b}\eta^{a}-\eta^{b}\na_{b}\xi^{a}=0,\nn\\
\pounds_{\zeta}\eta^{a}&=&\zeta^{b}\na_{b}\eta^{a}-\eta^{b}\na_{b}\zeta^{a}=0,\nn\\
\pounds_{\xi}\zeta^{a}&=&\xi^{b}\na_{b}\zeta^{a}-\zeta^{b}\na_{b}\xi^{a}=0.\label{poundxizeta}
\eea

Now we find the first variation as follows~\cite{scherk75} \bea
\frac{dS}{d\alpha} &=&\int\int d\tau
d\sigma~\eta_{b}(\xi^{a}\na_{a}P_{\tau}^{b}+\zeta^{a}\na_{a}P_{\sigma}^{b})\nn\\
& &-\int d\sigma~P_{\tau}^{b}\eta_{b}|_{\tau=0}^{\tau=T}-\int
d\tau~P_{\sigma}^{b}\eta_{b}|_{\sigma=0}^{\sigma=2\pi},\label{dsdalpha}
\eea where the world sheet currents associated with $\tau$ and
$\sigma$ directions are respectively given by~\cite{scherk75} \bea
P_{\tau}^{a}&=&\frac{1}{f}[(\xi\cdot\zeta)\zeta^{a}-(\zeta\cdot\zeta)\xi^{a}],\nn\\
P_{\sigma}^{a}&=&\frac{1}{f}[(\xi\cdot\zeta)\xi^{a}-(\xi\cdot\xi)\zeta^{a}].\label{pps}\eea
Using the endpoint conditions $\eta^{a}(0)=\eta^{a}(T)=0$ and
periodic condition $\eta^{a}(\sigma+2\pi)=\eta^{a}(\sigma)$, we
have the geodesic surface equation~\cite{scherk75} \beq
\xi^{a}\na_{a}P_{\tau}^{b}+\zeta^{a}\na_{a}P_{\sigma}^{b}=0,\label{geodesic}\eeq
and the constraint identities~\cite{scherk75} \beq
\begin{array}{ll}
P_{\tau}\cdot\zeta=0, &P_{\tau}\cdot P_{\tau}+\zeta\cdot\zeta=0,\\
P_{\sigma}\cdot\xi=0, &P_{\sigma}\cdot P_{\sigma}+\xi\cdot\xi=0.
\end{array}
\label{consts} \eeq

Let $\gamma_{\alpha}(\tau,\sigma)$ denote a smooth one-parameter
family of geodesic surfaces: for each $\alpha\in {\mathbf R}$, the
tube $\gamma_{\alpha}$ is a geodesic surface parameterized by
affine parameters $\tau$ and $\sigma$.  For an infinitesimally
nearby geodesic surface in the family, we then have the following
geodesic surface deviation equation \bea
&&\xi^{b}\na_{b}(\eta^{c}\na_{c}P_{\tau}^{a})+\zeta^{b}\na_{b}(\eta^{c}\na_{c}P_{\sigma}^{a})
\nn\\
&&+R_{bcd}^{~~~a}(\xi^{b}P_{\tau}^{d}+\zeta^{b}P_{\sigma}^{d})\eta^{c}\equiv
(\Lambda\eta)^{a}=0. \eea For a small variation $\eta^{a}$, our
goal is to compare $S(\alpha)$ with $S(0)$ of the string.  The
second variation $d^{2}S/d\alpha^{2}(0)$ is then needed only when
$dS/d\alpha (0)=0$.  Explicitly, the second variation is given by
\bea & &\frac{d^{2}S}{d\alpha^{2}}|_{\alpha =0}=-\int\int d\tau
d\sigma~\left[(\eta^{c}\na_{c}P_{\tau}^{b})(\xi^{a}\na_{a}\eta_{b})\right.\nn\\
&
&\left.+(\eta^{c}\na_{c}P_{\sigma}^{b})(\zeta^{a}\na_{a}\eta_{b})
-R_{acb}^{~~~d}(\xi^{a}P_{\tau}^{b}+\zeta^{a}P_{\sigma}^{b})\eta^{c}\eta_{d}\right]\nn\\
& &-\int
d\sigma~P_{\tau}^{b}\eta^{a}\na_{a}\eta_{b}|_{\tau=0}^{\tau=T}
-\int
d\tau~P_{\sigma}^{b}\eta^{a}\na_{a}\eta_{b}|_{\sigma=0}^{\sigma=2\pi}.\nn\\
\label{dds} \eea Here the boundary terms vanish for the fixed
endpoint and the periodic conditions, even though on the geodesic
surface we have breaks which we will explain later. After some
algebra using the geodesic surface deviation equation, we have
\beq \frac{d^{2}S}{d\alpha^{2}}|_{\alpha =0}=\int\int d\tau
d\sigma~\eta_{a}(\Lambda\eta)^{a}.\label{dds2}\eeq

\section{Jacobi fields in orthonormal gauge}
\setcounter{equation}{0}
\renewcommand{\theequation}{\arabic{section}.\arabic{equation}}

The string action and the corresponding equations of motion are
invariant under reparameterization
$\tilde{\sigma}=\tilde{\sigma}(\tau,\sigma)$ and
$\tilde{\tau}=\tilde{\tau}(\tau,\sigma)$.  We have then gauge
degrees of freedom so that we can choose the orthonormal gauge as
follows~\cite{scherk75} \beq
\xi\cdot\zeta=0,~~~\xi\cdot\xi+\zeta\cdot\zeta=0,\label{gauge}\eeq
where the plus sign in the second equation is due to the fact that
$\xi\cdot\xi$ is timelike and $\zeta\cdot\zeta$ is spacelike. Note
that the gauge fixing (\ref{gauge}) for the world sheet
coordinates means that the tangent vectors are orthonormal
everywhere up to a local scale factor~\cite{scherk75}.  In this
parameterization the world sheet currents (\ref{pps}) satisfying
the constraints (\ref{consts}) are of the form \beq
P_{\tau}^{a}=-\xi^{a},~~~P_{\sigma}^{a}=\zeta^{a}.\eeq The
geodesic surface equation and the geodesic surface deviation
equation read \beq
-\xi^{a}\na_{a}\xi^{b}+\zeta^{a}\na_{a}\zeta^{b}=0,\label{geodesic2}\eeq
and \bea
&&-\xi^{b}\na_{b}(\xi^{c}\na_{c}\eta^{a})+\zeta^{b}\na_{b}(\zeta^{c}\na_{c}\eta^{a})\nn\\
&&-R_{bcd}^{~~~a}(\xi^{b}\xi^{d}-\zeta^{b}\zeta^{d})\eta^{c}
=(\Lambda\eta)^{a}=0.\label{devieq2} \eea

We now restrict ourselves to strings on constant scalar curvature
manifold such as $S^{n}$.  We take an ansatz that on this manifold
the string shape on the geodesic surface $\gamma_{0}$ is the same
as that on a nearby geodesic surface $\gamma_{\alpha}$ at a given
time $\tau$. We can thus construct the variation vectors
$\eta^{a}(\tau)$ as vectors associated with the centers of the
string of the two nearby geodesic surfaces at the given time
$\tau$. We then introduce an orthonormal basis of spatial vectors
$e_{i}^{a}$ $(i=1,2,...,n-2 )$ orthogonal to $\xi^{a}$ and
$\zeta^{a}$ and parallelly propagated along the geodesic surface.
The geodesic surface deviation equation (\ref{devieq2}) then
yields for $i,j=1,2,...,n-2$ \beq
\frac{d^{2}\eta^{i}}{d\tau^{2}}+(R_{\tau j\tau}^{~~~i}-R_{\sigma
j\sigma}^{~~~i})\eta^{j}=0.\label{rrs} \eeq The value of
$\eta^{i}$ at time $\tau$ must depend linearly on the initial data
$\eta^{i}(0)$ and $\frac{d\eta^{i}}{d\tau}(0)$ at $\tau=0$. Since
by construction $\eta^{i}(0)=0$ for the family of geodesic
surfaces, we must have \beq
\eta^{i}(\tau)=A^{i}_{~j}(\tau)\frac{d\eta^{j}}{d\tau}(0).\label{etaitau}\eeq
Inserting (\ref{etaitau}) into (\ref{rrs}) we have the
differential equation for $A^{i}_{~j}(\tau)$ \beq
\frac{d^{2}A^{i}_{~j}}{d\tau^{2}}+(R_{\tau k\tau}^{~~~i}-R_{\sigma
k\sigma}^{~~~i})A^{k}_{~j}=0,\label{daij}\eeq with the initial
conditions \beq
A^{i}_{~j}(0)=0,~~~\frac{dA^{i}_{~j}}{d\tau}(0)=\delta^{i}_{~j}.\label{initials}\eeq
Note that in (\ref{daij}) we have the last term originated from
the contribution of string property.

Next we consider the second variation equation (\ref{dds}) under
the above restrictions \beq \frac{d^{2}S}{d\alpha^{2}}|_{\alpha
=0}=\int\int d\tau
d\sigma~\left(\frac{d\eta^{i}}{d\tau}\frac{\eta_{i}}{d\tau}
-(R_{\tau j\tau}^{~~~i}-R_{\sigma
j\sigma}^{~~~i})\eta^{j}\eta_{i}\right).\label{dds3} \eeq  We
define the index form $I_{\gamma}$ of a geodesic surface $\gamma$
as the unique symmetric bilinear form $I_{\gamma}:
T_{\gamma}\times T_{\gamma}\rightarrow {\mathbf R}$ such that \beq
I_{\gamma}(V,V)=\frac{d^{2}S}{d\alpha^{2}}|_{\alpha=0}\eeq for
$V\in T_{\gamma}$. From (\ref{dds3}) we can easily find \bea
I_{\gamma}(V,W)&=&\int\int d\tau
d\sigma~\left(\frac{dW^{m}}{d\tau}\frac{dV_{m}}{d\tau}\right.\nn\\
& &\left.-(R_{\tau j\tau}^{~~~m}-R_{\sigma
j\sigma}^{~~~m})W^{j}V_{m}\right).\label{igamma} \eea If we have
breaks $0=\tau_{0}<\cdots<\tau_{k+1}=T$, and the restriction of
$\gamma$ to each set $[\tau_{i-1},\tau_{i}]$ is smooth, then the
tube $\gamma$ is piecewise smooth.  The variation vector field $V$
of $\gamma$ is always piecewise smooth.  However $dV/d\tau$ will
generally have a discontinuity at each break $\tau_{i}$ $(1 \leq i
\leq k)$.  This discontinuity is measured by \beq
\Delta\frac{dV}{d\tau}(\tau_{i})=\frac{dV}{d\tau}(\tau_{i}^{+})
-\frac{dV}{d\tau}(\tau_{i}^{-}),\eeq where the first term derives
from the restrictions $\gamma|[\tau_{i},\tau_{i+1}]$ and the
second from $\gamma|[\tau_{i-1},\tau_{i}]$.  If $\gamma$ and $V\in
T_{\gamma}$ have the breaks $\tau_{1}<\cdots<\tau_{k}$, we have
\beq \sum_{i=0}^{k}\int_{\tau_{i}}^{\tau_{i+1}}\frac{d}{d\tau}
\left(V_{m}\frac{dW^{m}}{d\tau}\right)d\tau
=-\sum_{i=0}^{k}V_{m}\Delta\frac{dW^{m}}{d\tau}(\tau_{i}) \eeq to
yield \bea I_{\gamma}(V,W)&=&-\int\int d\tau
d\sigma~V^{m}\left(\frac{d^{2}W^{m}}{d\tau^{2}}\right.\\
&&\left.+(R_{\tau j\tau}^{~~~m}-R_{\sigma
j\sigma}^{~~~m})W^{j}\right)\nn\\
&&-\sum_{i=0}^{k}\int
d\sigma~V_{m}\Delta\frac{dW^{m}}{d\tau}(\tau_{i}).\label{igamma2}
\eea Here note that if we do not have the breaks, (\ref{dds3})
yields \beq \frac{d^{2}S}{d\alpha^{2}}|_{\alpha =0}=-\int\int
d\tau
d\sigma~\eta_{i}\left(\frac{d^{2}\eta^{i}}{d\tau^{2}}+(R_{\tau
j\tau}^{~~~i}-R_{\sigma
j\sigma}^{~~~i})\eta^{j}\right).\label{ddsnobr} \eeq

A solution $\eta^{a}$ of the geodesic surface deviation equation
(\ref{rrs}) is called a Jacobi field on the geodesic surface
$\gamma$. A pair of strings $p, q\subset\gamma$ defined by the
centers of the closed strings on the geodesic surface is then
conjugate if there exists a Jacobi field $\eta^{a}$ which is not
identically zero but vanishes at both strings $p$ and $q$.
Roughly speaking, $p$ and $q$ are conjugate if an infinitesimally
nearby geodesic surface intersects $\gamma$ at both $p$ and $q$.
From (\ref{etaitau}), $q$ will be conjugate to $p$ if and only if
there exists nontrivial initial data: $d\eta^{i}/d\tau (0)\neq 0$,
for which $\eta^{i}=0$ at $q$. This occurs if and only if $\det
A^{i}_{~j}=0$ at $q$, and thus $\det A^{i}_{~j}=0$ is the
necessary and sufficient condition for a conjugate string to $p$.
Note that between conjugate strings, we have $\det A^{i}_{~j}\neq
0$ and thus the inverse of $A^{i}_{~j}$ exists. Using (\ref{daij})
we can easily see that \beq
\frac{d}{d\tau}\left(\frac{dA_{ij}}{d\tau}A^{i}_{~k}-A_{ij}\frac{dA^{i}_{~k}}{d\tau}\right)=0.
\label{ddtauda}\eeq In addition, the quantity in parenthesis of
(\ref{ddtauda}) vanishes at $p$, since $A^{i}_{~j}(0)=0$. Along a
geodesic surface $\gamma$, we thus find \beq
\frac{dA_{ij}}{d\tau}A^{i}_{~k}-A_{ij}\frac{dA^{i}_{~k}}{d\tau}=0.\label{daijdtau}\eeq

If $\gamma$ is a geodesic surface with no string conjugate to $p$
between $p$ and $q$, then $A^{i}_{~j}$ defined above will be
nonsingular between $p$ and $q$.  We can then define
$Y^{i}=(A^{-1})^{i}_{~j}\eta^{j}$ or $\eta^{i}=A^{i}_{~j}Y^{j}$.
From (\ref{ddsnobr}) and (\ref{daijdtau}), we can easily verify
\beq \frac{d^{2}S}{d\alpha^{2}}|_{\alpha =0}=\int\int d\tau
d\sigma~\left(A_{ij}\frac{dY^{j}}{d\tau}\right)^{2}\geq 0. \eeq
Locally $\gamma$ minimizes the Nambu-Goto string action, if
$\gamma$ is a geodesic surface with no string conjugate to $p$
between $p$ and $q$.

On the other hand, if $\gamma$ is a geodesic surface but has a
conjugate string $r$ between strings $p$ and $q$, then we have a
non-zero Jacobi field $J^{i}$ along $\gamma$ which vanishes at $p$
and $r$. Extend $J^{i}$ to $q$ by putting it zero in $[r,q]$. Then
$dJ^{i}/d\tau(r^{-})\neq 0$, since $J^{i}$ is nonzero.  But
$dJ^{i}/d\tau(r^{+})=0$ to yield \beq
\Delta\frac{dJ^{i}}{d\tau}(r)=-\frac{dJ^{i}}{d\tau}(r^{-})\neq
0.\eeq We choose any $k^{i}\in T_{\gamma}$ such that \beq
k_{i}\Delta\frac{dJ^{i}}{d\tau}(r)=c,\label{kidel}\eeq with a
positive constant $c$. Let $\eta^{i}$ be $\eta^{i}=\epsilon
k^{i}+\epsilon^{-1}J^{i}$ where $\epsilon$ is some constant, then
we have \beq
I_{\gamma}(\eta,\eta)=\epsilon^{2}I_{\gamma}(k,k)+2I_{\gamma}(k,J)+
\epsilon^{-2}I_{\gamma}(J,J).\label{epsilons}\eeq By taking
$\epsilon$ small enough, the first term in (\ref{epsilons})
vanishes and the third term also vanishes due to the definition of
the Jacobi field and (\ref{igamma2}).  Substituting (\ref{kidel})
into (\ref{igamma2}) we have $I_{\gamma}(k,J)=-2\pi c$ and thus
\beq \frac{d^{2}S}{d\alpha^{2}}|_{\alpha =0}=-4\pi c,\eeq which is
negative definite.  From the above arguments, we conclude that
given a smooth timelike tube $\gamma$ connecting two strings $p,
q\subset M$, the necessary and sufficient condition that $\gamma$
locally minimizes the surface of the closed string tube between
$p$ and $q$ over smooth one parameter variations is that $\gamma$
is a geodesic surface with no string conjugate to $p$ between $p$
and $q$. It is also interesting to see that on $S^{n}$, the first
non-minimal geodesic surface has $n-1$ conjugate strings as in the
case of point particle.  Moreover, on the Riemannian manifold with
the constant sectional curvature $K$, the geodesic surfaces have
no conjugate strings for $K<0$ or $K=0$, while conjugate strings
occur for $K>0$~\cite{cheeger75}.

\section{Conclusions}
\setcounter{equation}{0}
\renewcommand{\theequation}{\arabic{section}.\arabic{equation}}

The Nambu-Goto string action has been introduced to study the
geodesic surface equation in terms of the world sheet currents
associated with $\tau$ and $\sigma$ directions.  By constructing
the second variation of the surface spanned by closed strings, the
geodesic surface deviation equation has been discussed for the
closed strings on the curved manifold.

Exploiting the orthonormal gauge, the index form of a geodesic
surface has been defined together with breaks on the string tubes.
The geodesic surface deviation equation in this orthonormal gauge
has been derived to find the Jacobi field on the geodesic surface.
Given a smooth timelike tube connecting two strings on the
manifold, the condition that the tube locally minimizes the
surface of the closed string tube between the two strings over
smooth one parameter variations has been also discussed in terms
of the conjugate strings on the geodesic surface.

In the Morse theoretic approach to the string theory, one could
consider the physical implications associated with geodesic
surface congruences and their expansion, shear and twist. It would
be also desirable if the string topology and the Gromov-Witten
invariant can be investigated by exploiting the Morse theoretic
techniques.  These works are in progress and will be reported
elsewhere.

\acknowledgments The work of YSC was supported by the Korea
Research Council of Fundamental Science and Technology (KRCF),
Grant No. C-RESEARCH-2006-11-NIMS, and the work of STH was
supported by the Korea Research Foundation (MOEHRD), Grant No.
KRF-2006-331-C00071, and by the Korea Research Council of
Fundamental Science and Technology (KRCF), Grant No.
C-RESEARCH-2006-11-NIMS.


\begin{thebibliography}{99}
\bibitem{witten87} M.B. Green, J.H. Schwarz and E. Witten, Superstring Theory Vol. 1
(Cambridge Univ. Press, Cambridge, 1987).
\bibitem{pol98} J. Polchinski, String Theory Vol. 1 (Cambridge Univ. Press, Cambridge, 1999).
\bibitem{witten82} E. Witten, J. Diff. Geom. {\bf 17}, 661 (1982).
\bibitem{morse34} M. Morse, The Calculus of Variations in the Large
(Amer. Math. Soc., New York, 1934).
\bibitem{milnor63} J. Milnor, Morse Theory (Princeton Univ. Press, Princeton, 1963).
\bibitem{wald84} R.M. Wald, General Relativity (The Univ. of Chicago Press, Chicago, 1984).
\bibitem{floer} A. Floer, Comm. Pure Appl. Math. {\bf 41}, 393 (1988).
\bibitem{schwarz} M. Schwarz, Morse Homology, Vol. 111 of Prog. Math. (Birkh\"auser,
Basel, 1993).
\bibitem{majer} A. Abbondandolo, P. Majer, Comm. Pure
Appl. Math. {\bf 54}, 689 (2001).
\bibitem{chas99} M. Chas and D. Sullivan, String Topology, to appear in Ann. Math., {\tt math.GT/9911159}.
\bibitem{biran04} P. Biran, O. Cornea and F. Lalonde, Morse Theoretic Methods in Nonlinear Analysis and
in Symplectic Topology Series II: Mathematics, Physics and
Chemistry, Vol. 217 of NATO Sci. Series (Springer, New York,
2004).
\bibitem{mcduff94} D. McDuff and D. Salamon, J-holomorphic Curves and Quantum Cohomology,
Vol. 6 of Univ. Lecture Series (Amer. Math. Soc., Providence,
1994).
\bibitem{hong03} S.T. Hong, J. Geom. Phys. {\bf 48}, 135 (2003).
\bibitem{atiyah75} M.F. Atiyah, V. Patodi and I. Singer, Math. Proc. Camb. Phil. Soc. {\bf 77}, 43 (1975);
Math. Proc. Camb. Phil. Soc. {\bf 78}, 405 (1975); Math. Proc.
Camb. Phil. Soc. {\bf 79}, 71 (1976).
\bibitem{witten89} E. Witten, Comm. Math. Phys. {\bf 121}, 351 (1989).
\bibitem{nambu70} Y. Nambu, Lecture at the
Copenhagen Symposium, 1970, unpublished; T. Goto, Prog. Theor.
Phys. {\bf 46}, 1560 (1971).
\bibitem{scherk75} J. Scherk, Rev. Mod. Phys. {\bf 47}, 123
(1975); J. Govaerts, Lectures given at Escuela Avanzada de Verano
en Fisica, Mexico City, Mexico (1986).
\bibitem{cheeger75} J. Cheeger and D. Ebin, Comparison
Theorems in Riemannian Geometry (North-Holland, Amsterdam, 1975).

\end{thebibliography}
\end{document}